# Why Are You More Engaged? Predicting Social Engagement from Word Use

| Jalal Mahmud | Jilin Chen | Jeffrey Nichols |
| IBM Research - Almaden | IBM Research - Almaden | IBM Research - Almaden |
| 650 Harry Rd, San Jose | 650 Harry Rd, San Jose | 650 Harry Rd, San Jose |
| jumahmud@us.ibm.com | jilinc@us.ibm.com | jwnichols@us.ibm.com |

**ABSTRACT**
We present a study to analyze how word use can predict social engagement behaviors such as replies and retweets in Twitter. We compute psycholinguistic category scores from word usage, and investigate how people with different scores exhibited different reply and retweet behaviors on Twitter. We also found psycholinguistic categories that show significant correlations with such social engagement behaviors. In addition, we have built predictive models of replies and retweets from such psycholinguistic category based features. Our experiments using a real world dataset collected from Twitter validates that such predictions can be done with reasonable accuracy.

**ACM Classification Keywords**
H.5.2 [Information Interfaces and Presentation]: User Interfaces - Interaction styles.

**INTRODUCTION**
Recent years have seen a rapid growth in micro-blogging and the rise of popular micro-blogging services, such as Twitter where millions of tweets are posted every day. There have been a number of research works that have explored users' tweeting activity, such as replies and retweets [12, 15, 2]. Twitter has also been used to engage with strangers [13, 10]. Recently Chen et al. has investigated users' engagement behavior in Twitter relating to a political movement [6]. Understanding factors of users' social engagement can benefit social question answering [13, 10, 12, 15], viral marketing [3], and information diffusion [1], etc.

In this paper, we focus on understanding how users' word usage may affect different social engagement behaviors on Twitter. Prior research reported correlations of people's word usage with personality, and predicting personality from word usage [20, 7, 11]. In addition, algorithms that use word-based features to predict other attributes, such as sentiment [14] and political polarization [4] also exist. However, word use has not been studied in the context of different types of social engagement behaviors, such as response and retweets in Twitter.

In this work, we attempt to understand the role of word usage in two social engagement behaviors on Twitter: replies and retweets. We measured word use in a number of psycholinguistic categories as defined by the Linguistic Inquiry and Word Count (LIWC) dictionary [16]. Then, we correlated such psycholinguistic category scores with replies and retweets. Based on our analysis with a real-world dataset collected from Twitter, we have found statistically significant correlations of several psycholinguistic categories with replies and retweets. Furthermore, we have built predictive models from psycholinguistic category features for predicting likelihood of such replies and retweets on Twitter.

Our work contributes to both theory and practice. On the theoretical side, our work is the first study that reports correlations of word usage with social engagement behaviors such as replies and retweets. On the practical side, our prediction models for predicting likelihood of replies and retweets from word usage can be used in wide varieties of scenarios such as recommending potential answerers for social question-answering [10, 12, 15], or potential information propagators for spreading alerts and warning messages in an emergency [17], viral marketing [3] and campaigns ranging from politics and government to social issues. Our experiments demonstrate that replies and retweets can be predicted reasonably accurately (23.6-29.9% Mean Absolute Error and 76-85% binary classification accuracy). This result shows the promise of developing predictive models for wide varieties of social behaviors (e.g., following, liking, mentions) from word use and using such models in personalized systems.

**DATASETS**
We used Twitter's Streaming API to randomly sample 17640 users in July 2013. Then we used Twitter's REST

API to obtain their past 200 tweets. We computed each user's *past_response_rate*, which is the ratio of each user's number of responses to the total number of questions asked on Twitter. We used a rule-based method of using only question (?) mark to detect questions. Our choice is justified by previous studies which have found that most (81.5% as reported in [12]) questions asked on social network sites contained a question (?) mark. Also a rule-based method of using only question (?) mark to identify question in online content achieves more than 97% precision in detecting questions [4]. We did not consider retweets and tweets containing URLs. The average past_response_rate was 0.754 and standard deviation was 0.097.

| LIWC categories | Pearson Correlation | Significance Level |
|---|---|---|
| Anger | -0.173 | ** |
| Cognition | 0.152 | ** |
| Communication | 0.163 | * |
| Anxiety | -0.083 | * |
| Social Process | 0.104 | * |
| Positive Feelings | 0.125 | * |

Table 1. Statistically significant correlations (* means p < 0.05, ** means p < 0.01) of LIWC categories with response

We also collected each user's previous retweet history and computed each user's *past_retweet_rate*, which is the ratio of the number of retweets the user made to the total number of user's tweets. The average past_retweet rate was 0.117 and standard deviation was 0.15.

**PSYCHO-LINGUISTIC ANALYSIS FROM TEXT**

We measured word uses with the Linguistic Inquiry and Word Count (LIWC) 2001 dictionary [16]. LIWC is the most commonly used language analysis tool for investigating the relation between word use and psychological variables [16]. LIWC 2001 defines over 70 different categories, each of which contains several dozens to hundreds of words [16]. We excluded the categories that are non-semantic (e.g., proportion of long words) or relevant to speech (e.g. fillers), and considered the remaining 66 LIWC categories.

For each person, we computed his/her LIWC-based scores in each category as the ratio of the number of occurrences of words in that category in one's tweets and the total number of words in his/her tweets. We excluded retweets when computing LIWC-based personality scores, because retweets are content generated by others.

**RESPONSE ANALYSIS**

We begin by running a Pearson correlation analysis between the LIWC categories and users *past_response_rate*. The statistically significant correlations are shown in Table 1. LIWC category anger is negatively correlated with response which indicates that people who use more angry words (such as *anger*, *angry*) are less likely to respond. The LIWC category communication is significantly positively correlated with response, which indicates that more communicative people are more likely to respond. An anxious person, exhibited by the LIWC category anxiety and consisting of words such as *afraid* and *alarm*, is less likely to respond. The LIWC category cognition, exemplified by words such as *accept*, *acknowledge*, *admit*, and *agree*, has a significant positive correlation with response. Similarly, a social (exemplified by words such as *interact, involve*), and individual with positive feelings (exemplified by words such as *care*, *cheer*, *attachment*) are more likely to respond. These are quite intuitive.

| LIWC categories | Correlation | Significance Level |
|---|---|---|
| Perception | 0.251 | ** |
| Communication | 0.144 | ** |
| Social Process | 0.145 | * |
| Physical States | -0.172 | * |
| Tentative | -0.053 | * |
| Positive Feelings | 0.193 | * |
| Positive Emotions | 0.067 | * |
| Inclusive | 0.21 | * |
| Other Refs | 0.175 | ** |

Table 2. Statistically significant correlations (* means p < 0.05, ** means p < 0.01) of LIWC categories with retweet

**RETWEET ANALYSIS**

We also ran a Pearson correlation analysis between the LIWC categories and *past_retweet_rate*. The significant correlations are shown in Table 2. The LIWC categories perception, communication, social process, positive feelings, positive emotions, inclusive and other refs show positive significant correlations with retweeting behavior. The LIWC categories communication, social process and positive feelings also exhibited similar characteristics with response behavior.

So a person who is more communicative, social and has positive feelings is more responsive and more willing to retweet. The LIWC category perception represents words such as *ask*, *call*, and *contact*. These words are often used by people who are more interactive, and it seems intuitive that a person who scores high in the perception category is also more likely to retweet. The LIWC category positive emotions is exemplified by words such as *accept* and *admit*.

| Independent variables used | Mean Absolute Error (MAE) | |
|---|---|---|
| | Response prediction | Retweet prediction |
| All LIWC categories | 0.299 | 0.297 |
| LIWC categories with significant correlations | 0.253 | 0.236 |

**Table 3. Regression results over a 10 fold cross-validation**

This is quite intuitive that a person who scores high in such category is more likely to retweet. In addition, we found that a person who scores high in the LIWC category inclusive, containing words such as *along* and *also*, is also more likely to retweet. This category is intuitively related with people who are more social and friendly, and it follows that people who scores high also seem to retweet. A person who scores high in the LIWC category "other refs," exemplified by words such as *he*, *she*, and *they*, is also more likely to retweet others' tweets.

The LIWC category "physical states" has significant negative correlations with retweet behavior. This category represents words such as *diabetes*, *disease*, *dizziness*, and *sleep*. These words often indicate someone's sickness or inactivity. This may indicate apathy on the part of the user, which would make them less likely to retweet. However, we lack a good explanation of the significant negative correlation of LIWC category tentative (represents words such as *luck*, *may*, *perhaps*) and retweet behavior.

**PREDICTION MODELS**

We attempted to build predictive models for response and retweet based on the psycholinguistic category features to understand their predictive power. We performed both regression analysis and a classification study using WEKA [19], a widely used machine learning toolkit.

For regression analysis, we formulated linear regressions to predict response and retweet rates using LIWC measures. Thus, LIWC attributes were independent variables of the regression model and users' *past_response_rate/ past_retweet_rate* was the dependent variable. We tried a number of regression approaches, including simple linear regression, multiple linear regression, pace regression, logistic regression and SVM regression. We performed 10-fold cross validation. SVM regression slightly outperformed other algorithms.

Table 3 presents the result of the regression analysis in terms of mean absolute error (MAE). We find that response can be predicted within 25.3%-29.9% MAE and retweet can be predicted within 23.6%-29.7% MAE. The best result is obtained when LIWC categories with significant correlations were used as independent variables in the regression model.

In the classification study, we used supervised binary machine learning algorithms to classify users with above-median levels of *past_response_rate/past_retweet_rate*. We experimented with a number of classifiers from WEKA [19], including naive Bayes, SVM, J48 (a decision-tree-based classifier), Random Forest (an ensemble method that combines multiple decision trees).

Table 4 shows the classification result of the best WEKA classifier in terms of AUC under 10-fold cross validation. We see that classifying high (above median) or low (below median) responders/retweeters from LIWC category based features can be done quite accurately. The best classification result is obtained when LIWC categories with significant correlations are used as features for the classification model.

| Features Used | Area Under ROC Curve (AUC) | |
|---|---|---|
| | Response prediction | Retweet prediction |
| All LIWC categories | 0.757 | 0.723 |
| LIWC categories with significant correlations | 0.801 | 0.849 |

**Table 4. Classification results over a 10-fold cross validation**

**DISCUSSION**

Our findings suggest several implications for social media interactions and personalized systems. Our work is the first study that reports correlations of word usage with different social engagements such as responses and retweets. Our work also discovers a set of psycholinguistic categories which have significant correlations with users' responses and retweets in Twitter. Such psycholinguistic categories are also shown to be useful to predict the likelihood of responses and retweets. Our findings also suggest that certain word use can predict such likelihoods. For instance, people who use angry words (such as *anger, hate*, *kill*) are less likely to respond, people who write about communication are more likely to respond, and people who write about positive feelings are more likely to retweet.

Predictive models of response behavior from word usage can identify users who are more likely to respond to a question. Question-asking systems such as [12, 13] can use the recommendations/ranks produced from such predictions to send questions to potential answerers. This can increase response rates of such systems [12, 13]. Predictive models of retweet from word uses can identify users who are more willing to retweet a tweet. This can increase the rate of retweeting which is useful for various information propagation applications including viral marketing [3], propagating emergency news [17] and social/government/political campaigns (e.g., a campaign to support "vaccination"). For example, when anti-government messages are spread in social media, government would want to spread counter messages to balance that effort and hence identify users who are more likely to retweet those counter messages. Our research to predict retweet behavior from word uses and identify potential retweeters is orthogonal to existing research on influence modeling and information diffusion [1] for the above applications.

Our regression analysis confirms that such predictions can be done with reasonably low error (23.6-29.9% MAE). We also observed comparable prediction errors for predicting responses and retweets. Furthermore, our classification study shows that classifying high/low responders/retweeters can be done quite accurately (76-85% binary classification accuracy). In this study we did not explore ranking algorithms. However, learning-to-rank algorithms [9] may be useful to further harness the predictive information from word use.

Based on this finding, we hope that other social behaviors such as mentions and follow behavior on Twitter or liking on Facebook may be predictable from word uses. Predicting such interactions from word uses can be useful for followee recommendations [8] or recommending users to mention [18]. Moreover, we may explore word based activity prediction (such as usage of specific hashtag) for followers of a specific account within the context of a campaign [6]. In addition, a word-based predictive model for response can be useful to predict the potential responders in a question-asking forum.

**CONCLUSIONS**

In this paper, we have presented a study on understanding how people with different word usage exhibit different social engagement behaviors through an analysis of responses and retweets on Twitter. We conducted psycho-linguistic analysis from word usage and found certain psycholinguistic categories have significant correlations with such social engagement behavior. In addition, our findings suggest that a few psycholinguistic categories share common characteristics for both responding and retweeting behavior. We have built predictive models of social engagement behaviors from the psycho-linguistic category features and demonstrated that our models are reasonably accurate to predict such social engagement behaviors.

In future, we would like to extend our analytics to other social media platforms, verify the generality of the findings with other forms of social engagement, and continue deeper investigation of predictive models for social engagement based on word uses.